# High Mobility Free-Standing InSb Nanoflags Grown On InP Nanowire Stems For Quantum Devices

*Isha Verma[1], Sedighe Salimian[1], Valentina Zannier[1*], Stefan Heun[1], Francesca Rossi[2], Daniele Ercolani[1*], Fabio Beltram[1], and Lucia Sorba[1]*

[1] NEST, Istituto Nanoscienze-CNR and Scuola Normale Superiore, Piazza San Silvestro 12, I-56127 Pisa, Italy

[2] IMEM-CNR, Parco Area delle Scienze 37/A, I-43124 Parma, Italy

**Abstract.** High quality heteroepitaxial two-dimensional (2D) InSb layers are very difficult to realize owing to the large lattice mismatch with other widespread semiconductor substrates. A way around this problem is to grow free-standing 2D InSb nanostructures on nanowire (NW) stems, thanks to the capability of NWs to efficiently relax elastic strain along the sidewalls when lattice-mismatched semiconductor systems are integrated. In this work, we optimize the

[1]*correspondence email: valentina.zannier@nano.cnr.it , daniele.ercolani@sns.it

morphology of free-standing 2D InSb nanoflags (NFs). In particular, robust NW stems, optimized growth parameters, and the use of reflection high-energy electron diffraction (RHEED), to precisely orient the substrate for preferential growth, are implemented to increase the lateral size of the 2D InSb NFs. Transmission electron microscopy (TEM) analysis of these NFs reveals defect-free zinc blend crystal structure, stoichiometric composition, and relaxed lattice parameters. The resulting NFs are large enough to fabricate Hall-bar contacts with suitable length-to-width ratio enabling precise electrical characterization. An electron mobility of ~29,500 $cm^2$/Vs is measured, which is the highest value reported for free-standing 2D InSb nanostrutures in literature. We envision the use of 2D InSb NFs for fabrication of advanced quantum devices.

**Introduction**

High-quality III-V narrow band-gap semiconductor materials with strong spin-orbit coupling and large Landé g-factor provide a promising platform for applications in the field of optoelectronics, spintronics, and quantum computing. Indium antimonide (InSb) offers a narrow band gap, high carrier mobility and a small effective mass, and perfectly fits to this scope. In fact, it has attracted tremendous attention in recent years, both theoretically[1,2] and experimentally[3-6], for the implementation of topological superconducting states supporting Majorana zero modes (MZMs). In particular, high quality InSb nanowires (NWs) have opened new research arenas in quantum transport, since their geometry leads to carrier confinement and their electron energy levels are electrostatically tunable. However, the challenge remains, as NW morphology does not provide enough flexibility to fabricate multi-contact Hall-bar devices. An alternative geometry, that would allow a high degree of freedom in device fabrication and give the opportunity to explore new material properties, is represented by two-dimensional (2D) InSb nanostructures.

Unfortunately, owing to large lattice mismatch between InSb and other widespread semiconductor systems, the growth of high-quality heteroepitaxial 2D InSb layers is complicated and demands stacks of buffer layers. To counter this problem, one can grow free-standing 2D InSb nanostructures on NW stems, thanks to the limited size of the heterointerface and the capability to efficiently relax elastic strain along the sidewalls. Nevertheless, controlling the aspect ratio of free-standing InSb nanostructures is challenging, due to the low vapor pressure of Sb and the surfactant effect.[7] In general, the narrower growth window of III-Sb in comparison to other III-Vs (III-P and III-As) is due to the surfactant effect of Sb atoms, as the atoms tend to segregate to the surface, thereby modifying the surface energy.[7,8] For this reason, it is essential to investigate the growth mechanisms and the morphology of InSb free-standing nanostructures.

Studies on the synthesis and the characterization of free-standing 2D InSb nanostructures grown by catalyst-assisted molecular beam epitaxy (MBE) and metal-organic vapor phase epitaxy (MOVPE) are present in literature[9-11]. Pure zinc-blend (ZB) InSb nanosheets were grown by Ag-assisted MBE.[9] In [10] and [11], a twin plane boundary induced the formation of 2D InSb nanosails/nanoflakes. Here, we have realized single-crystal free-standing InSb nanoflags (NFs) via Au-assisted chemical beam epitaxy (CBE).

In our previous work, we reported on the growth and morphology control of InSb nanostructures such as nanocubes, nanowires, and nanoflags on top of InAs NW stems[12]. Concerning the 2D shape, we found that the limitation in achieving InSb NFs larger than 1 µm in length and 280 nm in width was the flexibility of the thin untapered InAs NW stems. As the growth time of the asymmetric InSb segment is increased, the InAs stem bends, leading to the loss of the alignment with the precursor fluxes and consequently of the InSb orientation. Therefore, the preferential growth direction vanishes, and 3D-like InSb structures are obtained. A possible solution to avoid this problem is to employ more robust stems, as for example tapered NWs, to provide a more stable support for the InSb NFs. This will keep them well aligned even after long growth durations. Tapered InAs NWs are not so easy to obtain because of their wurtzite (WZ) crystal structure that reduces the radial vapor-solid growth on the NW sidewalls[13]. Instead, it is known that ZB or mixed WZ/ZB structures in NWs enhance the radial growth rate[13]. InP NWs grown by Au-assisted CBE on InP(111)B substrates have a mixed WZ/ZB crystal structure and a tapered morphology[14].

Keeping this into account, here we present the growth of InSb NFs on tapered and robust InP NWs. By using this approach, we are able to achieve (2.8 ± 0.2) µm long, (470 ± 80) nm wide, and (105 ± 20) nm thick InSb NFs. Furthermore, we have employed reflection high-energy electron diffraction (RHEED), in order to carefully adjust the substrate orientation with respect to the precursor beam fluxes, which minimizes the thickness of these NFs. Thanks to the larger dimension of the InSb NFs that we have obtained, we were able to realize Hall-bar contacts far enough to keep the standard length-to-with ratio between longitudinal and transversal contacts, which avoided the presence of mixed components in Hall-bar measurements and allowed to accurately investigate the electrical properties. We demonstrate an electron Hall mobility of ~29,500 $cm^2$/Vs at 4.2 K, which to our knowledge is the highest value reported so far for free-standing 2D InSb nanostructures [9-11].

**Experimental details**

The InP-InSb axial heterostructures of the present study were synthesized by CBE in a Riber Compact-21 system on InP(111)B substrates via Au-assisted growth[15,16] following a methodology very similar to that reported in [12]. We used 30 nm Au colloids dropcasted onto the bare substrate as seeds to catalyze the growth, and trimethylindium (TMIn), tert-butylphosphine (TBP), and trimethylantimony (TMSb) as metal-organic (MO) precursors.

The growth sequence of the InP-InSb heterostructures consists in the growth of InP stems followed by the InSb segments. We grew InP stems for 60 min at a growth temperature ($T_{InP}$) of 400°C using 0.6 Torr and 1.2 Torr of TMIn and TBP line pressures, respectively. Afterwards, the substrate temperature was ramped down by $\Delta T$, in presence of TBP flux only, to the InSb growth temperature, $T_{InSb}=T_{InP}+\Delta T$ ($\Delta T$ is negative here). In order to initiate InSb growth, group V flux was abruptly switched from TBP to TMSb. For the growth of the InSb segment, we used TMIn line pressures in the 0.3 – 0.9 Torr range and TMSb line pressures in the 0.8 – 2.4 Torr range, as described in the following section. The growth temperatures were measured with a pyrometer with overall accuracy of ±5°C. At the end of the growth, samples were cooled down to room temperature in an ultra-high vacuum (UHV) environment, without group V precursor flux, in order to prevent the accumulation of Sb on the heterostructure sidewalls.

In this work, we first demonstrate the optimization of InSb NWs on InP NW stems. From the knowledge of the effect of the growth parameters we progress towards InSb NF growth. The growth procedures of both InSb NWs and NFs are discussed in detail later in the text.

The InP NW stems and InSb NWs were grown rotating the substrate at 5 rotations per minute (rpm) for the whole growth time. Conversely, there was no sample rotation during the growth of the InSb NFs, and the orientation was carefully adjusted before starting their growth with the help of the RHEED pattern. Indeed, by stopping the rotation, we trigger asymmetric growth, which is crucial to achieve the NF morphology[12]. The alignment protocol is illustrated in Figure 1. InP NWs grown on InP(111)B have a hexagonal cross section with six equivalent {112} sidewalls, as visible from the SEM image (45°-tilted and top-view) of a representative NW (Fig. 1(a)). Panel (b) shows a schematic view of the InP NW inside the growth chamber (top- and side-view) with respect to the RHEED beam (red arrow). Before initiating the InSb NF growth, the <110> direction was identified using the RHEED pattern as illustrated in Fig. 1(c). Indeed this is the direction at which we can see the overlap of the WZ and ZB reciprocal lattices on the RHEED screen[17]. The substrate was then rotated by 30° to obtain an alignment with the precursor beam that maximizes the InSb NF elongation, as described in the "results and discussion" section, and the InSb NF growth was initiated.

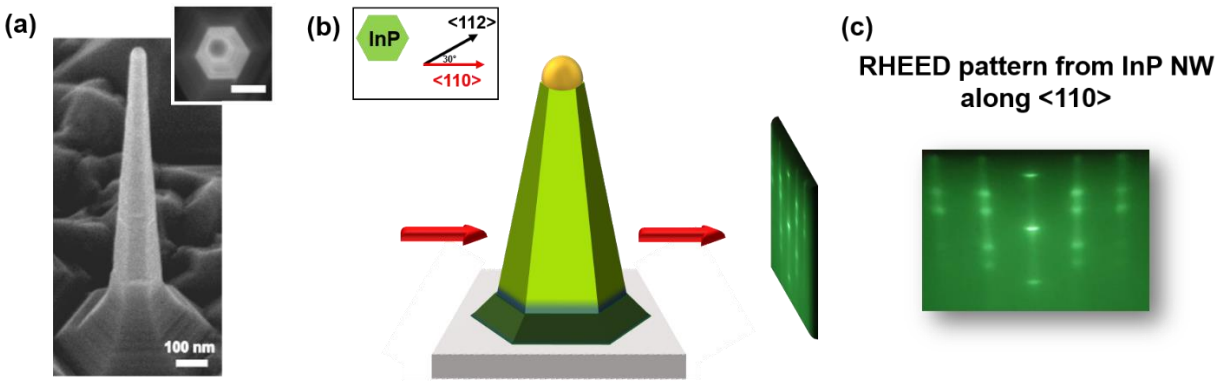

**Figure 1.** (a) 45°-tilted and top view (inset) SEM images of an InP NW stem (scale bar: 100 nm). (b) top view (inset) and side view representation of the alignment procedure of InP NWs with their corresponding RHEED pattern along the <110> direction (red arrow). (c) RHEED pattern of mixed WZ/ZB InP NWs in <110> direction.

The morphological characterization of the grown heterostructures was performed by acquiring field emission scanning electron microscopy (SEM) images with a Zeiss Merlin SEM operating at an accelerating voltage of 5 keV. Characterization of crystal structure and chemical composition of mechanically detached NFs was carried out by transmission electron microscopy (TEM) in a JEOL JEM-2200FS operated at 200 keV, equipped with an in-column $\Omega$ filter and Oxford X-ray energy dispersive spectrometer (EDS/EDX). Imaging was performed in high-resolution (HR) TEM mode combined with zero-loss energy filtering and by high angle annular dark-field in scanning mode (HAADF-STEM).

To fabricate the devices, the as-grown InSb NFs were dry transferred on a pre-patterned p-type Si(111) substrate, which serves as a global back gate. A 285 nm-thick $SiO_2$ layer covers the Si substrate. During the mechanical transfer, the InSb NFs are detached from the InP NW stems, so that well isolated InSb NFs were found lying randomly distributed on the substrate. Then the position of selected InSb NFs was determined relative to predefined alignment markers using SEM images. Considering the thickness and the edge geometry of the InSb NFs, electrodes were patterned on a 400 nm-thick layer of AR-P 679.04 resist with standard electron-beam lithography (EBL). Prior to metal deposition, the samples were chemically etched for 1 min in a 1:9 $(NH_4)_2S_x$ DI water-diluted solution at 40°C, to remove the native oxide layer from the exposed NF areas, and then rinsed for 30 s in $H_2O$. Next, a 10/190 nm Ti/Au film was deposited using thermal evaporation, followed by lift off. All low-temperature magneto-transport data that we present in this paper are from one single device, but we obtained consistent data from three other devices.

**Results and discussion**

*InSb NWs*

We first studied the effect of InSb growth temperature on the final shape of the InP-InSb heterostructured NWs. Figure 2 shows SEM images of 3 samples grown at different ΔT. For all samples, the growth of the InP NW stems was followed by InSb growth for 30 min using 0.6 Torr of TMIn and 1.2 Torr of TMSb, with sample rotation at 5 rpm for the whole growth duration.

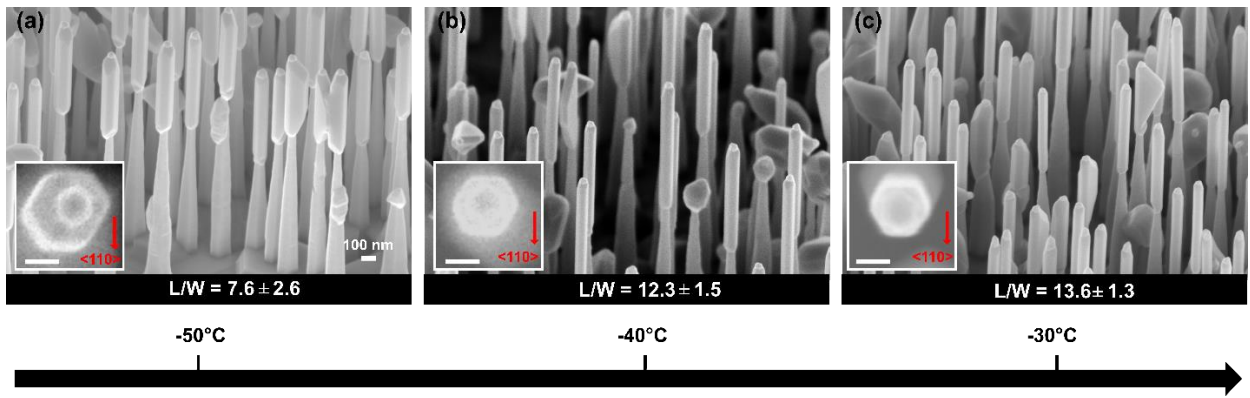

**Figure 2.** InP-InSb heterostructured NWs at different ΔT. 45°-titled SEM images of NWs obtained at (a) ΔT=-50°C, (b) -40°C and (c) -30°C (scale of all panels is the same as in (a)). Insets represent respective high magnification top view SEM images of one representative NW with red arrow indicating the <110> substrate direction (inset scale bars: 50 nm). The aspect ratio, L/W, of the InSb NWs is denoted for each ΔT at the bottom of the corresponding SEM image.

The aspect ratio, i.e length/width (L/W) of the InSb segments, is reported at the bottom of each panel. Larger values of ΔT, corresponding to lower InSb growth temperatures, enhance the InSb radial growth rate (larger diameter) and lower the axial growth rate, all together decreasing L/W. Increasing the temperature above ΔT=-30°C leads instead to InSb desorption, as in fact the InSb sublimation temperature is known to be around 400°C[18]. Therefore, ΔT=-30°C is the optimal InSb growth temperature to obtain high aspect ratio InSb NWs on top of InP NW stems. The insets in each panel of Fig. 2 show top view SEM images of an individual InP-InSb heterostructured NW with hexagonal cross section of the upper InSb segment comprising of six equivalent {110} oriented sidewalls.

To evaluate other growth parameters affecting the growth and the morphology of the InP-InSb heterostructured NWs, we grew the InSb segments at different TMIn/TMSb line pressure ratios.

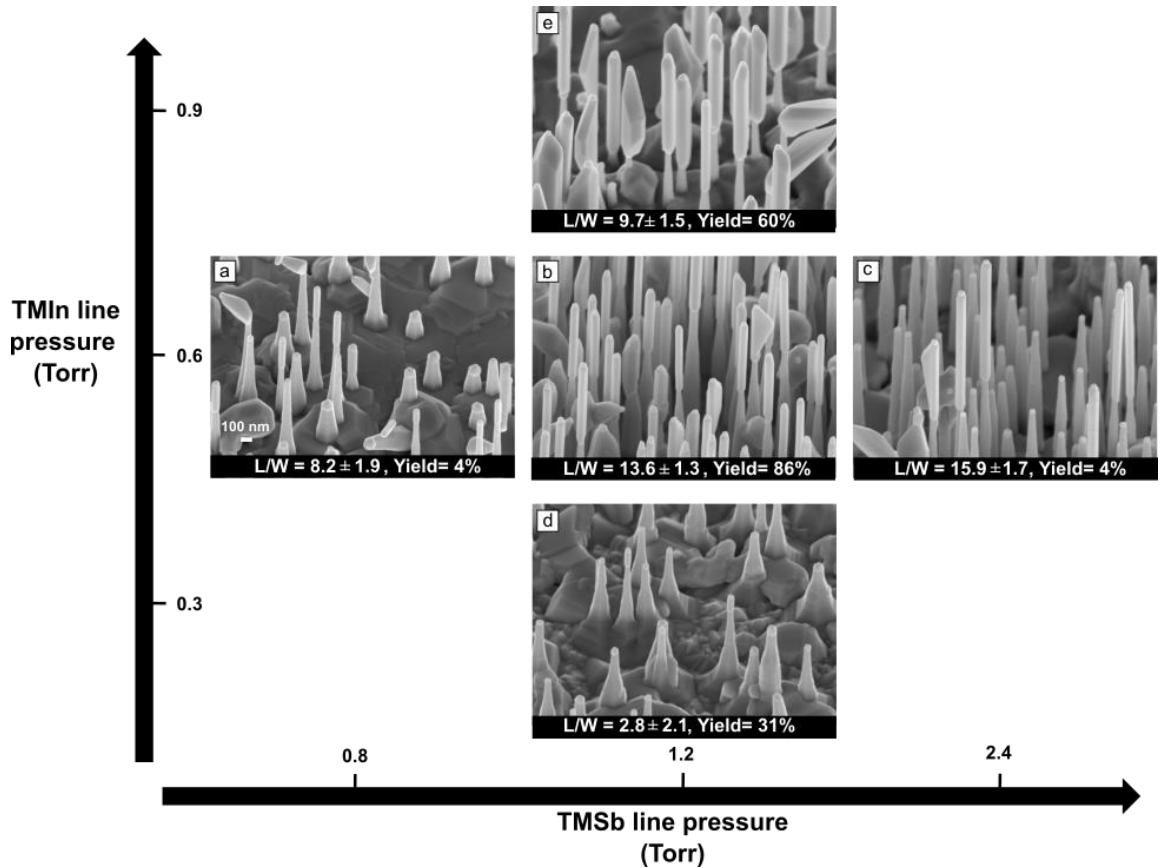

**Figure 3.** Yield and morphology map of InP-InSb heterostructured NWs as a function of the TMIn and TMSb line pressures. The InSb segments are grown in two configurations: constant TMIn and constant TMSb line pressure. (a-c) 45°-tilted SEM images of InSb segments for constant TMIn line pressure of 0.6 Torr and TMSb line pressure of (a) 0.8 Torr, (b) 1.2 Torr, and (c) 2.4 Torr. (d-e) 45°-tilted SEM images of InSb segments for constant TMSb line pressure of 1.2 Torr and TMIn line pressure of (d) 0.3 Torr and (e) 0.9 Torr. All images have the same scale indicated in (a).

Figure 3 illustrates the morphology of the InP-InSb heterostructured NWs, obtained as a function of TMIn and TMSb line pressure as employed for the InSb segment growth. The x- and y-axes denote TMSb and TMIn line pressures, respectively. All samples are grown at the optimal growth temperature of ΔT=-30°C. We find that the yield (i.e. the ratio between straight InSb nanostructures and total number of InP-InSb heterostructures - counting all straight, kinked and non-nucleating InSb) and the heterostructure morphology strongly depend on the precursor line pressures. The InSb segments are grown in two configurations: constant TMIn and varying TMSb, or constant TMSb but varying TMIn line pressure. For the series with constant TMIn line pressure (0.6 Torr) we grew 3 samples with TMSb line pressure of (a) 0.8 Torr, (b) 1.2 Torr, and (c) 2.4 Torr. We found that increasing Sb flux increases the aspect ratio of the NWs. The yield is 4% for both the samples grown at lower (TMSb=0.8 Torr) and at higher (TMSb=2.4 Torr) Sb flux. Instead, the yield is much higher (around 86%) for TMSb line pressure of 1.2 Torr.

Panels (d) and (e) of Fig. 3 show 45°-tilted SEM images of the InP-InSb heterostructured NWs obtained at constant TMSb line pressure of 1.2 Torr and TMIn line pressure of 0.3 Torr and 0.9 Torr, respectively. For constant TMSb line pressure, the highest L/W is obtained for TMIn/TMSb = 0.6/1.2. The InSb growth yield first increases from 31% to 86% by increasing the TMIn line pressure from 0.3 Torr to 0.6 Torr and then drops to 60% for 0.9 Torr of TMIn. Based on the MO line pressure experiment, we found that the maximum yield is obtained at TMIn/TMSb = (0.6/1.2) while the highest L/W of 15.9 is obtained at TMIn/TMSb= (0.6/2.4). Based on these results, we can conclude that the best conditions, at ΔT= -30°C, to obtain both high L/W and good yield for InSb NWs are TMIn line pressure of 0.6 Torr and TMSb line pressure in the range of 1.2 - 2.4 Torr.

It is worth noting that some elongated structures similar to flags are occasionally observed in the samples. However, these are very few (their occurrence is always < 15%) and randomly oriented objects among many NWs with symmetric cross section. It might be that these asymmetric structures are formed due to partial shadowing of the beam fluxes by neighboring NWs, or to some other local effects that we did not study in detail.

*InSb NFs*

To grow free-standing InSb NFs, we can exploit our knowledge derived from the growth optimization of InSb NW on InP stems (as discussed above) and from the asymmetric InSb growth and its elongation via substrate orientation and higher Sb flux (from previous work reported in [12]). InP NW stems were grown with sample rotation for 90 min to provide more sturdy support, followed by InSb growth at $\Delta T=-30°C$ without rotation, with abrupt switch in group V flux from TBP to TMSb, without variation in the TMIn flux. The growth protocol employed for the growth of InSb NFs is schematically shown in panel (a) of Figure 4. The initial InSb growth comprises of 0.6 Torr of TMIn and 2.3 Torr of TMSb to have a high Sb flux but still a good yield, and then additional 60 min of growth, linearly increasing the TMSb line pressure from 2.3 Torr to 2.6 Torr. Such Sb flux grading helps to enhance the asymmetric growth, increasing the lateral dimensions of the NFs, without compromising too much the yield of the InSb growth on top of the InP NW stems that is known to drop if the growth starts directly with higher Sb flux (see Fig. 3).

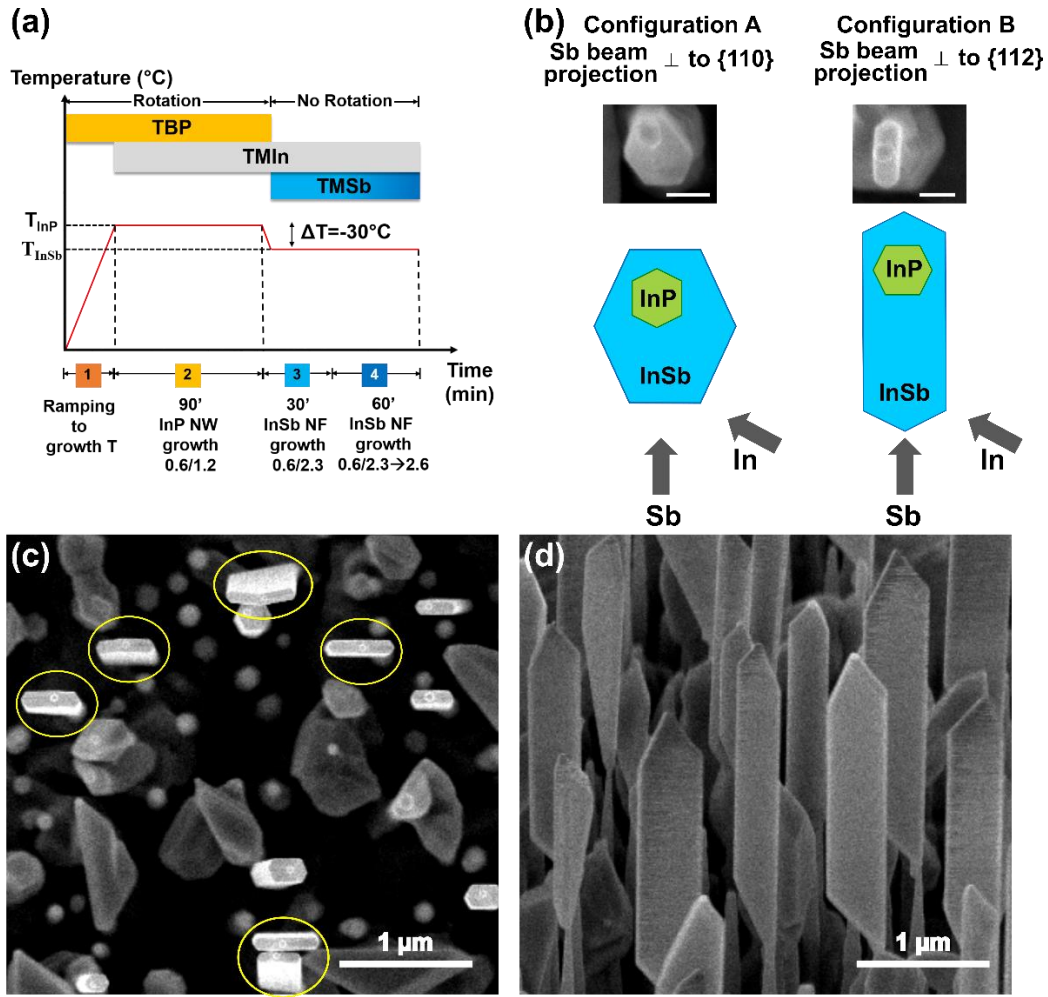

**Figure 4.** Growth and orientation protocol of InSb NFs. (a) Schematics of the growth protocol developed for obtaining InSb NFs. (b) Top-view schematics of the precursor beams projection with respect to the InP-InSb heterostructure cross section (bottom) and corresponding SEM images after the first 30 minutes of InSb growth (top) in the two possible configurations: configuration A with the Sb beam projection perpendicular to a {110} plane, and configuration B with the Sb beam projection perpendicular to a {112} plane. The scale bar is 100nm. (c) Top view and (d) 45°-tilted SEM image of the InSb NFs obtained in configuration B (scale bar: 1 μm). The InSb NFs which have (w/t) ≥ 4 are marked by yellow circles in panel (c).

As mentioned in the experimental details, before initiating the growth of InSb NFs, the substrate rotation was stopped, and the orientation of the NWs was carefully adjusted with the help of the RHEED pattern. Orientation choices are illustrated in panel (b) of Fig. 4: the InP NW (represented in green) exhibits a hexagonal cross section with six equivalent {112} sidewalls. The InSb segments (shown in blue) still have a hexagonal cross section, but with six equivalent {110} sidewalls. A schematic top view of InP-InSb heterostructures is shown for two orientations: configuration A and B. In configuration A, the Sb beam flux projection is perpendicular to a {110} sidewall of the InSb NWs, while in configuration B, the Sb beam projection is perpendicular to a {112} InP NW sidewall. We found that the growth in configuration A leads to thicker InSb NFs, while the growth in configuration B results in thinner NFs for the same growth time. This is explained by considering that the growth of the InSb segment involves two growth mechanisms that simultaneously occur: the VLS axial growth on top of the NW stem and the VS radial growth that is enhanced by high Sb flux[12]. If the sample was rotated, the radial growth would be uniform on the six {110} facets, and we would obtain InSb nanowires with symmetric cross section, showing six sidewalls equivalent in width. Conversely, when we stop the sample rotation and align the substrate in configuration B, there are only two {110} facets facing the Sb injector, i.e reached by direct impingement, so the growth rate on these two facets will be higher compared to the other four sidewalls, and we obtain thinner flags. On the other hand, when the sample is oriented in configuration A, only the three backside InSb facets (opposite to the Sb beam) are totally screened from Sb impingement, while the sidewall perpendicular to the Sb beam projection will receive the direct beam, and the two adjacent inclined facets will be reached by the beam at grazing incidence. Therefore, the NFs will be larger and less elongated. So, once we found the <110> direction with the help of the RHEED pattern at the end of the InP NWs growth (as shown in Fig. 1), we rotated the substrate by 30° (i.e. to configuration B) and started the InSb growth.

Panels (c) and (d) of Fig. 4 show top view and 45°-tilted SEM images of a sample grown in configuration B. The yield of the NFs, i.e nanostructures showing a width-to-thickness ratio (w/t) ≥ 4 (marked with yellow circles in panel (c)) is 40% and they have average length, width and thickness of (2.8 ± 0.2) μm, (470 ± 80) nm and (105 ± 20) nm, respectively. With the appropriate parameters (growth temperature and precursor fluxes), robust InP stems, and precise substrate orientation, we could grow the InSb NFs for longer time, obtaining larger InSb NFs with similar thickness, compared with the InSb NFs obtained on InAs NW stems[12] (see also Supporting Information section S1), which are suitable for making electronic devices, as demonstrated later. In addition, we speculate that similar morphologies might be achieved by adopting this directional growth approach for other materials if they show a consistent radial growth together with the axial elongation.

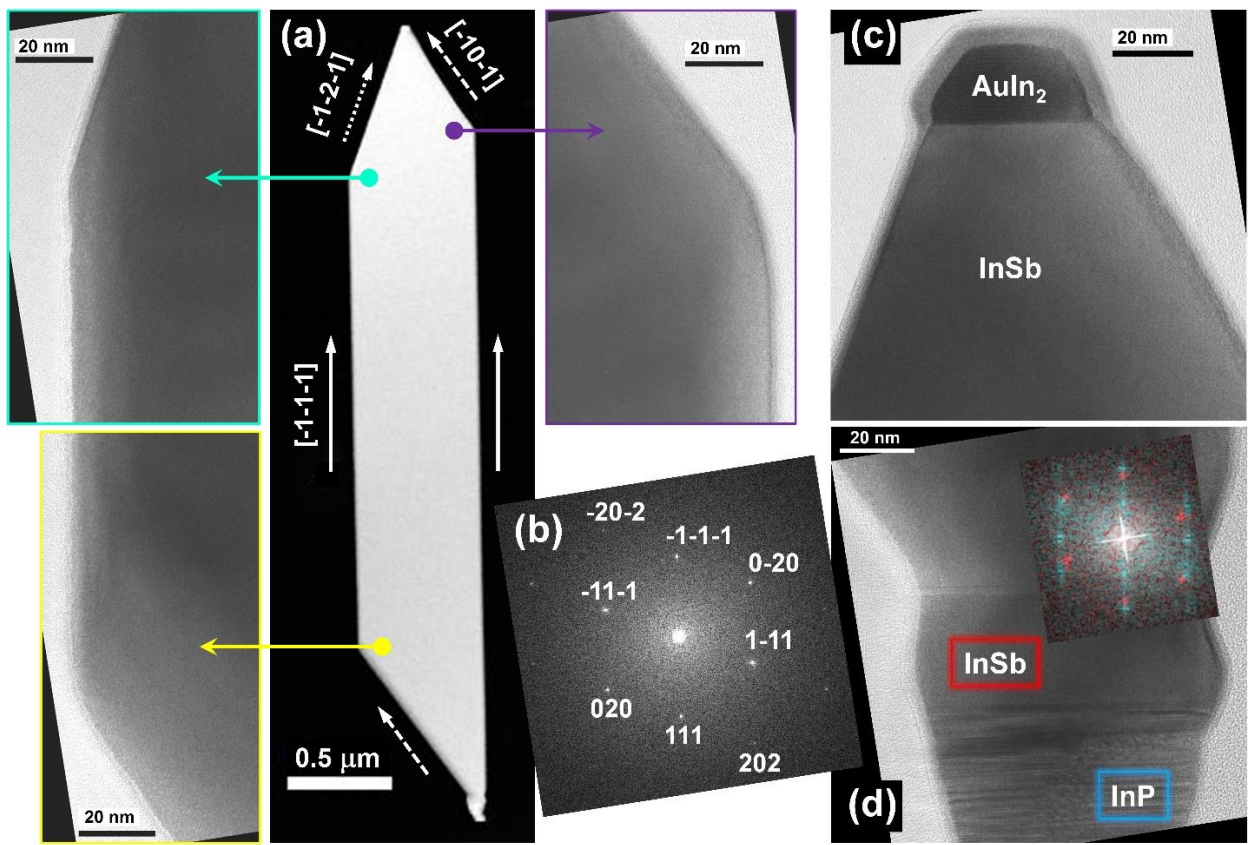

**Figure 5.** (a) STEM-HAADF image and corresponding HRTEM images acquired at the NF corners, in [10-1] zone axis; (b) indexed FFT of the HRTEM image framed in yellow in panel (a). (c) and (d) HRTEM images acquired at NF tip and base, respectively. The inset to (d) shows the FFT obtained by color mixing the FFTs of a small square region in InP (blue color) and InSb (red).

To determine the crystal quality of the NFs, their structure was analyzed by TEM (Figure 5). A STEM-HAADF overview of a single InSb NF, with a short segment of its InP stem and the catalyst particle at the tip, is shown in Fig. 5(a). The corresponding HRTEM images acquired at the three NF corners (purple-, green- and yellow-framed panels) show the defect-free InSb zinc blende lattice. The lattice spacing and the interplanar angles (see also the Fast Fourier Transform in Fig. 5(b)) match those of relaxed zinc blende InSb (JCPDS card 6-208). The analysis of the NFs faceting confirms the indexing observed in our previously grown samples described in detail in [12]. The major flat facets are of the type (10-1) and (-101), bordered by sides parallel to 3 pairs of directions: [-10-1] (dashed arrows), [-1-2-1] (dotted arrows) and [-1-1-1] (solid arrows, aligned with the growth axis).

At the NF tip, a sharp interface between InSb and the metal alloy seed particle is observed (Fig. 5(c)). EDX spectroscopy performed in STEM spot mode on several NFs allowed to identify the metal alloy components as Au and In and to quantify an atom gold content of 34±2%, consistent with $AuIn_2$.

At the NF base, both an axial and a radial growth of InSb on InP is observed, as shown in Fig. 5(d) and in Figure S2 of the Supporting Information. As clearly seen in these HRTEM images, the InP crystal structure is highly defected, with a mixed WZ/ZB stacking. Indeed, the energetic differences for hexagonal or cubic stacking sequences in the ⟨111⟩ direction are very small[19]. As a consequence, stacking faults easily occur in InP NWs vertically grown on (111)B substrates, resulting in NWs with alternating WZ/ZB segments[14]. The radially grown InSb, which was observed to be either asymmetric or symmetric around the InP stem, as shown in Figure S2 of the Supporting Information from panel (a) to (c), also appears defected, showing several stacking faults and twins in its ZB lattice along the stem length. On the contrary, the axially-grown InSb shows such a defected structure only in its initial layers, but after the first 10 nm the stacking becomes regular and a perfect ZB structure is recovered. After that, only a twin is observed occasionally within the first 50 nm (as shown in Figure S2 (a) and (c) of the Supporting Information). FFT analysis performed on the axial InSb close to the InP interface (Fig. 5(d), indicated in red with respect to the blue InP) shows that it reaches a complete relaxation. EDX maps (as shown in Figure S2 of the Supporting Information) confirm the purity (50 at% each for Indium and Antimony) and homogeneity of InSb.

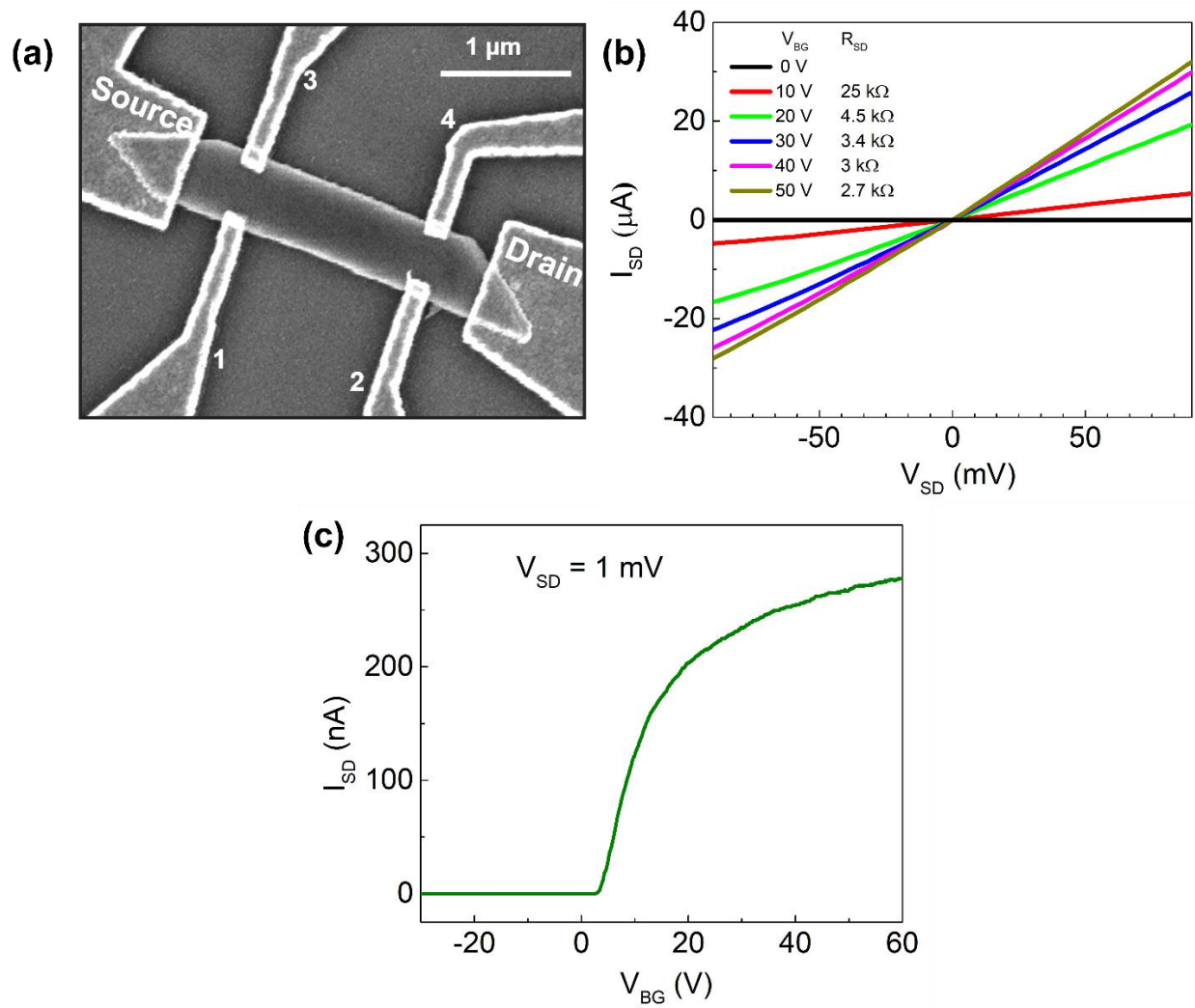

**Figure 6.** (a) SEM image of an InSb NF Hall-bar device with corresponding numbers for Hall-bar contacts. (b) Two-probe $I_{SD}$-$V_{SD}$ curves as a function of back gate voltage $V_{BG}$. (c) Source-drain current vs. back gate voltage, measured under 1 mV constant AC voltage bias. All measurements are performed at a temperature of 4.2 K.

To investigate the electronic properties of the InSb NFs, we performed low-temperature (4.2 K) magnetoresistance measurements on Hall-bar devices. A SEM image of a representative Hall-bar device is shown in Fig. 6(a). Figure 6(b) shows current–voltage ($I_{SD}$-$V_{SD}$) curves of the Source-Drain (S-D) channel at 4.2 K as a function of back gate voltage $V_{BG}$. The linear $I_{SD}$–$V_{SD}$ curves together with the low resistance values indicate the presence of good Ohmic contacts between the InSb NFs and the metal contacts, and the absence of a Schottky barrier. Increasing the back gate voltage, the source-drain resistance $R_{SD} = V_{SD}/I_{SD}$ decreases from 25 kΩ for $V_{BG}$ = 10 V to 2.7 kΩ for $V_{BG}$ = 50 V. In a measurement under constant AC voltage bias of 1 mV at 4.2 K, the variation of the injected current as a function of back gate voltage was measured and is shown in Fig. 6(c). A voltage bias is necessary here since in the depletion region (negative gate voltages), the sample is insulating, and a constant current could not flow. In other words, in the range of back gate voltages that we explored, we did not observe ambipolar behavior. The longitudinal voltage drop $V_{xx}$ is measured simultaneously in a four-probe configuration. We performed back gate sweeps for both $V_{xx}$ contact combinations [(1-2) and (3-4)], and both showed consistent results. From these data, the four-probe field-effect mobility is calculated to be about 28000 cm$^2$/Vs (for details, see Figure S3 of the Supporting Information). The charge carrier modulation shows an increasing conductance with increasingly positive back gate voltage, consistent with an n-type behavior of the InSb NFs and in agreement with the data shown in Fig. 6(b).

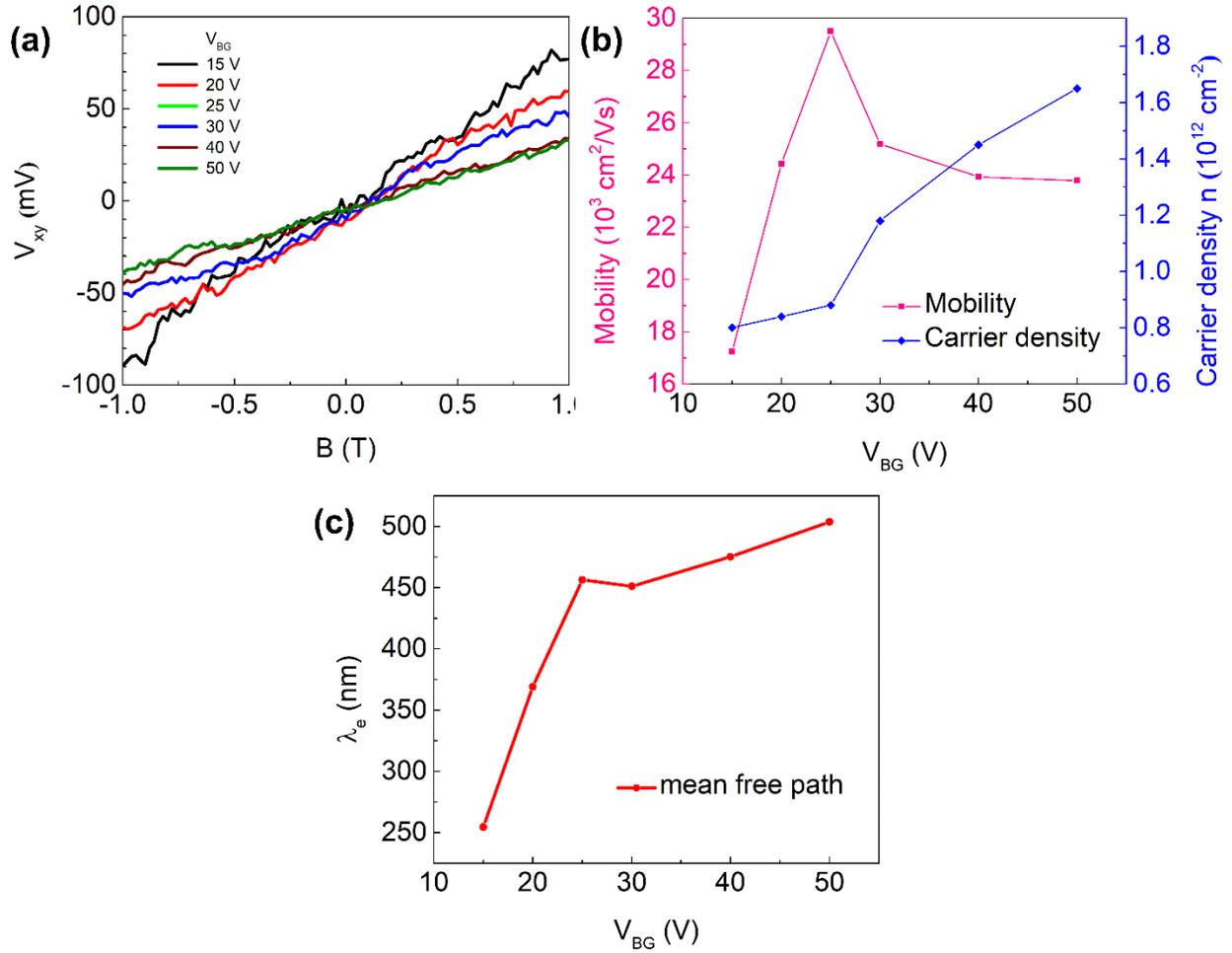

**Figure 7.** (a) Hall measurements on InSb NFs: $V_{xy}$ as a function of magnetic field B for different back gate voltages $V_{BG}$ at 4.2 K. (b) Mobility and charge carrier density obtained from the Hall measurements shown in (a). (c) Elastic mean free path $\lambda_e$ as a function of back gate voltage $V_{BG}$.

In addition to field-effect measurements, we performed a series of Hall-effect measurements using a constant AC current bias of 100 nA at 4.2 K. Since these measurements were performed in current bias, they start at a back gate voltage of 15 V, at which the channel is already well open. Figure 7(a) shows the resulting Hall-voltage curves as a function of magnetic field for different back-gate voltages $V_{BG}$. The corresponding charge-carrier densities and Hall-mobilities for various back-gate voltages are shown in Fig. 7(b) (for details, see Figure S4 of the Supporting

Information). Hall mobility increases with increasing back-gate voltage and shows a maximum of about 29,500 cm$^2$/Vs at $V_{BG}$ = 25 V, with a corresponding electron density of 8.5 x 10$^{11}$ cm$^{-2}$. For even higher carrier concentrations, mobility slightly drops again due to additional carrier scattering induced by Coulomb interactions. Hence, Hall mobility is in good agreement with the four-probe field effect mobility, and higher than in previous studies[9-11], which reported at most 20000 cm$^2$/Vs. We attribute this higher mobility to the fact that our flags are slightly thicker (100 nm) than the flakes reported previously that ranged from 50 nm to 80 nm, which reduces the contribution of surface- and interface-scattering. Figure 7(b) also shows that charge-carrier density increases with increasing back-gate voltage, as expected. Furthermore, we estimated the electron mean free path $\lambda_e$, using $\lambda_e = (\hbar\mu/e)(2\pi n)^{1/2}$ [20], with $\hbar$ the reduced Planck's constant and $n$ the 2D electron density from the Hall measurements (cf. Fig. 7(b)). As shown in Fig. 7(c), $\lambda_e$ reaches values of ~500 nm for $V_{BG} \geq$ 25 V, which compares favorably with literature[10, 11, 21].

**Conclusions**

In conclusion, we have realized free-standing 2D InSb NFs on InP NW stems exhibiting the highest electron mobility compared to other similar 2D InSb nanostructures reported in literature. This was possible by carefully choosing a robust supportive stem, tapered InP NWs, by optimizing the growth parameters leading to the growth of InSb NWs with high yield and high aspect ratio, and by aligning the samples in the direction that maximizes the NF elongation keeping the NF thickness at a minimum. This strategy allowed us to obtain InSb NFs of (2.8 ± 0.2) μm length, (470 ± 80) nm width and (105 ± 20) nm thickness, with defect-free ZB crystal structure, stoichiometric composition, and relaxed lattice parameters. We strongly believe that these NFs can serve for realization of exotic bound states at the semiconductor interface with superconductors, paving the way for the development of topological quantum computation technologies.

**Supporting Information**

The Supporting Information is available free of charge at https://pubs.acs.org/xxx. S1. InSb NFs on InAs stem vs InP stem. S2. EDX and HRTEM analysis of the InP-InSb NF interface. S3. Four-probe field effect mobility. S4 Hall mobility (PDF)

**Acknowledgments**

This research activity was partially supported by the SUPERTOP project, QUANTERA ERA-NET Cofound in Quantum Technologies (H2020 grant No. 731473), and by the FET-OPEN project AndQC (H2020 grant No. 828948).

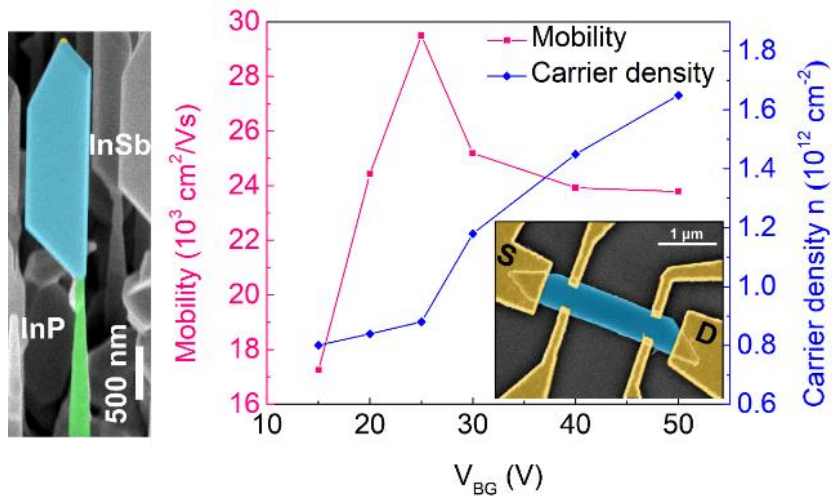

TOC graphics

Supporting Information

# "High Mobility Free-Standing InSb Nanoflags Grown On InP Nanowire Stems For Quantum Devices"

*Isha Verma[1], Sedighe Salimian[1], Valentina Zannier[1,\*], Stefan Heun[1], Francesca Rossi[2], Daniele Ercolani[1,\*], Fabio Beltram[1], and Lucia Sorba[1]*

[1] NEST, Istituto Nanoscienze-CNR and Scuola Normale Superiore, Piazza San Silvestro 12, I-56127 Pisa, Italy

[2] IMEM-CNR, Parco Area delle Scienze 37/A, I-43124 Parma, Italy

**Contents**

**S1. InSb NFs on InAs stem vs InP stem**

**S2. EDX and HRTEM analysis of the InP-InSb NF interface**

**S3. Four-probe field effect mobility**

**S4. Hall mobility**

[1]*correspondence email: valentina.zannier@nano.cnr.it , daniele.ercolani@sns.it

## S1. InSb NFs on InAs stem vs InP stem

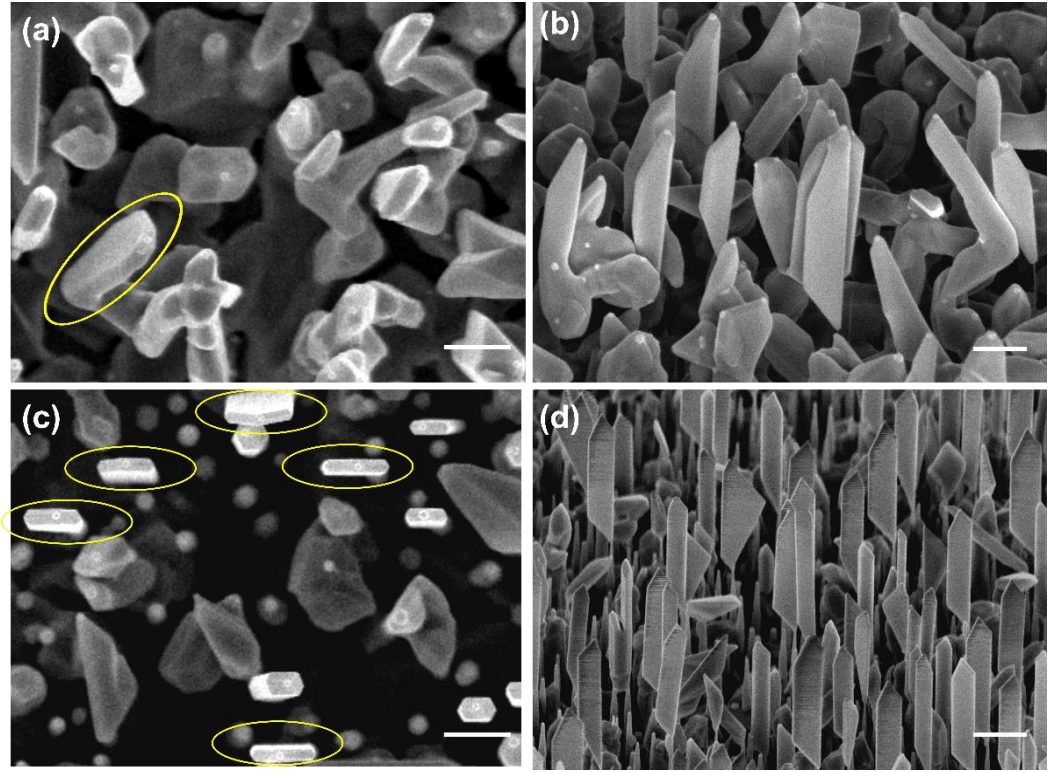

**Figure S1.** InSb NFs on InAs vs InP stem. Top view SEM images of InSb NFs on (a) InAs and (c) InP NW stems. 45°-tilted view SEM images of InSb NFs on (b) InAs stems and (d) InP NW stems. Scale bar is 500 nm in all panels. The InSb NFs that have (w/t) ≥ 4 are marked by yellow circles.

Figure S1 shows as-grown InSb NFs on InAs stems (top view and 45°-tilted view SEM image in panel (a) and (b), respectively), while on InP NW stems in panel (c) and (d). The growth protocol is the same as illustrated in panel (a) of Fig. 4 in the main text. The criterium for selection of preferred InSb NFs for fabricating quantum devices is defined by the following parameter:

$$\frac{\text{Width of InSb NFs}}{\text{Thickness of InSb NFs}} = \frac{w}{t} \geq 4$$

Counting the NFs satisfying this condition (marked with yellow circles in figure S1), we get a yield of 1% for InSb NFs grown on InAs stems, while for those grown on InP stems, the yield is 40%. This demonstrate a clear advantage for using robust tapered InP NW stems instead of thin untapered InAs stems. Indeed, for long InSb growth times, the thin untapered InAs stems bend, leading to the loss of alignment with the precursor fluxes and consequently of the InSb orientation. Therefore, the preferential growth direction vanishes, and more 3D-like InSb structures are obtained.

## S2. EDX and HRTEM analysis of the InP-InSb NF interface

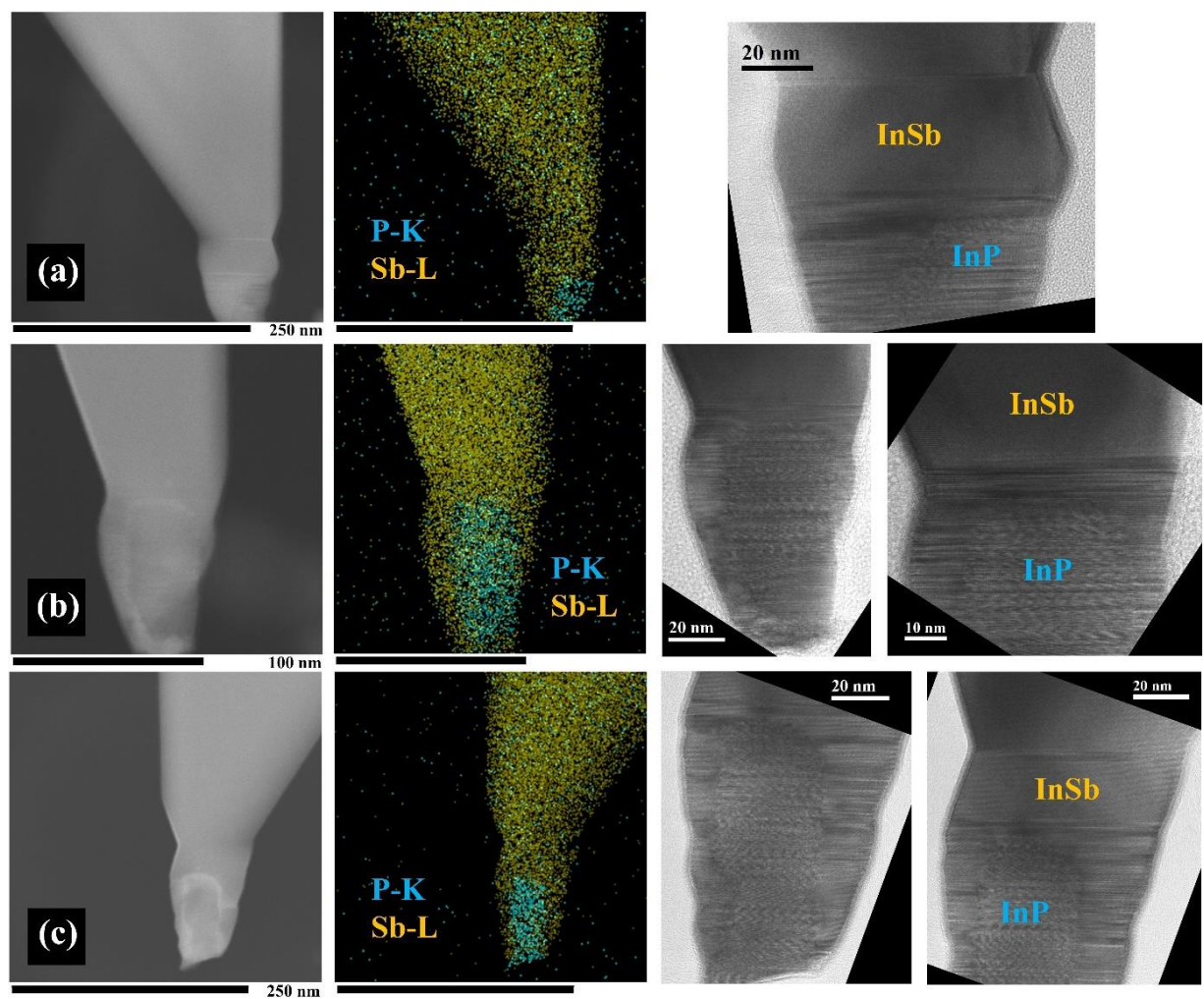

**Figure S2.** (a-c) EDX and HRTEM analysis of the basal region of three representative nanoflags, showing both axial and radial growth of InSb on the InP NW stem. From left to right: STEM-HAADF image, corresponding EDX map of the Sb (dark yellow) and P (blue) distributions, and HRTEM images of the interface.

Figure S2 shows STEM-EDX and HRTEM images of the basal region of three representative NFs, showing both axial and radial growth of InSb on the InP NW stem.

Since the as-grown InP NW stem has a mixed wurtzite/zinc blende structure, stacking defects are clearly visible. The InSb radial shell is grown in either an asymmetric or a symmetric fashion around the InP stem (see panels (a) to (c)) and it retains the stacking defects of the InP

NW stem (clearly visible from the HRTEM images). This radial InSb shell makes the top part of the InP NW stem thicker, adding further support for the growth of large InSb NFs.

## S3. Four-probe field effect mobility

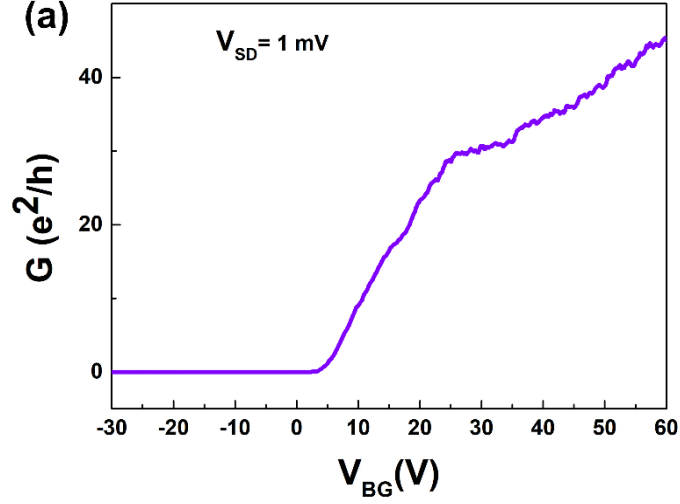

**Figure S3**. (a) Conductance G versus back gate voltage. The measurement was performed at a temperature of 4.2 K.

In a four-probe measurement under constant AC voltage bias at 4.2 K, the variation of the injected current is measured as shown in Figure 6(c) of the main text, and the longitudinal voltage drop $V_{xx}$ (using contacts 1-2) is measured, as well. This allows to calculate the conductance $G = I_{SD}/V_{xx}$ as shown in Figure S3(a). The four-probe field effect mobility is then obtained using the formula

$$\mu_{4p\ FE} = \frac{L}{WC_{ox}}\left(\frac{dG}{dV_{BG}}\right), \qquad (S1)$$

with L/W = 4.6 and $C_{ox}$ = 10 nF/cm$^2$. The four-probe field effect mobility obtained is 28000 cm$^2$/Vs. The n-type trend of the four-probe charge carrier modulation is in agreement with Fig. 6(c), since the conductivity increases with increasingly positive back gate voltages.

## S4. Hall mobility

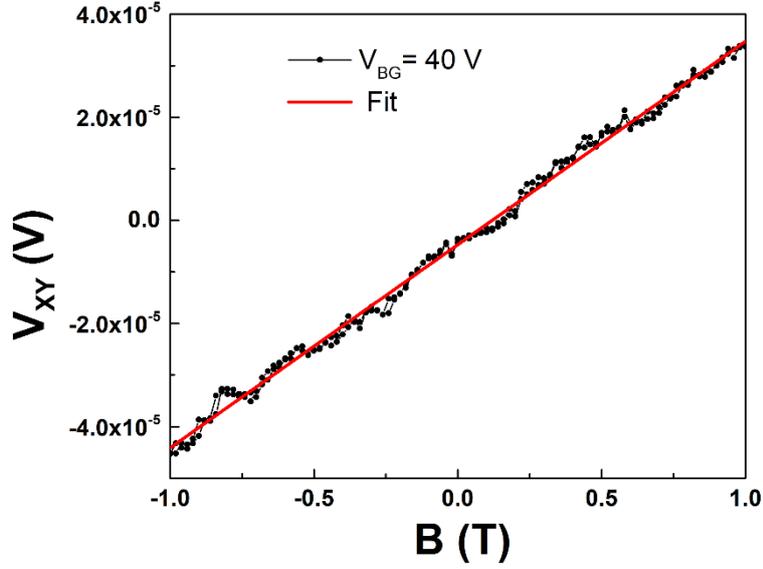

**Figure S4.** The transversal voltage drop $V_{xy}$ as a function of magnetic field B at $V_{BG}$ = 40 V and T = 4.2 K.

We measured the Hall voltages as a function of magnetic field ranging from -1T to +1T under constant AC current bias of 100 nA for different back gate voltages. We measured the Hall-bar device shown in Figure 6(a), with the channel width (width between contacts 1-3 and 2-4) of 325 nm and the channel length (1-2 and 3-4) of 1.5 μm. The naoflag thickness is ~100 nm. An example in Figure S4 shows $V_{xy}$ for $V_{BG}$ = 40 V. The figure shows the forward and backward sweep of the magnetic field, to demonstrate the reproducibility of the measurement, plus the fit to the experimental data, from which the carrier concentration is obtained. In detail, carrier concentration $n$ and Hall mobility $\mu_H$ for each back gate voltage are calculated using the formulas

$$\mu_H = \frac{L}{W<V_{xx}>}\left(\frac{V_{xy}}{B}\right) \qquad (S2)$$

$$n = \frac{I}{e}\left(\frac{B}{V_{xy}}\right) \qquad (S3)$$

with e the elementary charge.